\def\Roman#1{\uppercase\expandafter{\romannumeral#1}}
\documentclass[a4paper,12pt]{article}
\usepackage{cite}
\usepackage{amssymb,amsfonts}
\usepackage[mathscr]{eucal}
\usepackage{mathrsfs}
\usepackage[utf8x]{inputenc}

\title{Transition to the case of ``resolved gauge'' in 
the Lagrange-Poincar\'{e} equations for a mechanical system with symmetry on the total space of a principal fiber bundle whose base is    the bundle space of the associated bundle}

\author{S. N. Storchak}

\author{S. N. Storchak\\
\small{ A. A. Logunov Institute for High Energy Physics}\\
\small{of NRC ``Kurchatov Institute'',}\\
\small{Protvino, 142281, Russian Federation}}

\begin{document}

  \maketitle

\begin{abstract}
This note is a continuation of our earlier articles arXiv:1612.08897  and arXiv:1709.09030, where using the dependent coordinates the local Lagrange – Poincaré equations were obtained for a mechanical system with symmetry describing the motion of two interacting scalar particles on a special Riemannian manifold (the product of the total space of the principal fiber  bundle and vector space), on which a free proper and isometric action of a compact semisimple Lie group is given. Assuming the existence of the parametric representations for local sections in the principal bundle, we make  the transition to  independent coordinates in the obtained Lagrange- Poincaré  equations. 
\end{abstract}

In our previous papers \cite{Stor_1} and \cite{Stor_2}, we studied a mechanical system with symmetry describing the motion of two interacting scalar particles on a special Riemannian manifold: the product of the total space of the principal fiber bundle $ \rm P (\mathcal M, \mathcal G) $ and the vector space $V$. In addition, a free proper and isometric action of a compact semisimple Lie group on this manifold was also given.  From the general theory it follows that the configuration space of our system can be considered as the total space of the principal fiber bundle $\pi' : \mathcal P\times  V \to \mathcal P\times_{\mathcal G}  V $.   Applying the Poincaré variation principle to the system under consideration, we obtained the local Lagrange-Poincaré equations: 
\begin{equation}
N^A_B\frac{d{\omega}^B}{dt}+N^A_R\,{}^{\rm  H}\tilde {\Gamma}^R_{\tilde B\tilde M}{\omega}^{\tilde B} {\omega}^{\tilde M} +G^{EF}N^A_EN^{\tilde R}_F\Bigl[{ \mathscr F}^{\alpha}_{\tilde Q\tilde R}  {\omega}^{\tilde Q} p_{\alpha} +\frac12({\mathscr D}_{\tilde R} {d}^{\kappa\sigma}) p_{\kappa}p_{\sigma}+V_{,\tilde R}\Bigr]=0,                                 
\label{itog_hor_1}
\end{equation}
\begin{eqnarray}
&&\!\!\!\!\!\!\!\!\!\!\!\!N^r_B\frac{d{\omega}^B}{dt}+\frac{d{\omega}^r}{dt}+N^r_{\tilde R}\,{}^{\rm  H}\tilde {\Gamma}^{\tilde R}_{\tilde A\tilde B}{\omega}^{\tilde A} {\omega}^{\tilde B}+
G^{EF}N^r_FN^{\tilde R}_E\Bigl[{ \mathscr F}^{\alpha}_{\tilde Q\tilde R}  {\omega}^{\tilde Q} p_{\alpha}+
\nonumber\\
&&\!\!\!\!\!\!\!\!\!\!\!\!\frac12({\mathscr D}_{\tilde R} {d}^{\kappa\sigma}) p_{\kappa}p_{\sigma}+V_{,\tilde R}\Bigr]+
G^{rm}\Bigl[{ \mathscr F}^{\alpha}_{\tilde Q m}  {\omega}^{\tilde Q} p_{\alpha}+\frac12({\mathscr D}_{m} {d}^{\kappa\sigma}) p_{\kappa}p_{\sigma}+V_{,m}\Bigr]=0.
\label{itog_hor_2}
\end{eqnarray}

\begin{equation}
 \frac{d p_{\beta}}{dt}+ c^{\nu}_{\mu \beta}d^{\mu \sigma}p_{\sigma}p_{\nu}-c^{\nu}_{\sigma \beta}\mathscr A^{\sigma}_{\tilde E}\omega ^{\tilde E}p_{\nu}=0.  
\label{itog_vert}
\end{equation}
These equations were written in terms of dependent coordinates, which are typically used to describe dynamics in gauge theories. In our case, the dependent coordinates were implicitly determined by means of  equations representing the local sections  of the principal fiber bundle $ \rm P (\mathcal M, \mathcal G) $.

In this note it will be shown how  the Lagrange – Poincaré equations (\ref{itog_hor_1}), (\ref{itog_hor_2}) and (\ref{itog_vert})  are transformed into equations obtained under the assumption that for a local surface $ \Sigma $, which is a section in the principal bundle of $ \rm P (\mathcal M, \mathcal G) $, we know the parametric representation.\footnote{In physics, the case of the  parametric representation of such a local surface in the corresponding  principal fiber bundle is known as the case of gauges that can be resolved (the case of ``resolved gauges'').}

The use of such a surface makes it possible to describe the evolution on the base space $\mathcal M$ of $ \rm P (\mathcal M, \mathcal G) $ in terms of internal (invariant) coordinates. These coordinates are determined by means of  the invariant coordinate   functions $x^i(Q)$ given by the equations 
\[
 Q^{\ast}{}^A(x)=F^A(Q, a^{-1}(Q)).
\]
Also note, that on the local submanifold $\Sigma$, the coordinates $x^i$ satisfy the equations $\chi^{\alpha}(Q^{\ast}(x))=0$.

In our case, the local section $ \Sigma $ of $ \rm P (\mathcal M, \mathcal G) $ is used to determine the local section $\tilde \Sigma $ in the principal fiber bundle $\pi' : \mathcal P\times  V \to \mathcal P\times_{\mathcal G}  V $.
And the local surface $\tilde \Sigma$, in turn,  is necessary for the introduction of coordinates on the bundle $\pi' $.
That is, now, when we are given a parametric representation of $ \Sigma $, we can introduce  local coordinates
 $(x^i,\tilde f^b, a^{\alpha})$ instead of the coordinates $(Q^{\ast A},\tilde f^b, a^{\alpha})$. 

The transition to the new coordinates in the equations (\ref{itog_hor_1}), (\ref{itog_hor_2}) and (\ref{itog_vert}) is achieved by replacing the dependent coordinates $ Q ^{\ast A} $ with the functions $ Q^{\ast A} (x)$ in all terms of these equations.

First we consider   the horizontal equation (\ref {itog_hor_1}).

As a result of  the above replacement, the kinetic term $N^A_B\frac{d{\omega}^B}{dt}$ of the this equation becomes $Q^{\ast}{}^A_i
\ddot x^i+ N^A_B\,Q^{\ast}{}^A_{ij}\,\dot x^i \dot x^j$. This is due to the fact that ${\omega}^B=\frac{d Q^{\ast}{}^B}{dt}$, where  now $Q^{\ast}{}^B(t)\equiv Q^{\ast}{}^B(x(t))$, and also because     of  the equality $N^A_BQ^{\ast}{}^B_i=Q^{\ast}{}^A_i$. The latter follows from the representation $$ N^A_B(Q^{\ast}(x))= G^{\rm H}_{BR}(Q^{\ast}(x))Q^{\ast}{}^R_m h^{mn}Q^{\ast}{}^A_n,$$
where $G^{\rm H}_{BR}=G_{BR}-G_{BL}K^L_{\alpha}\gamma^{\alpha \beta}K^S_{\beta}G_{SR}$ and $h^{mn}$ is the inverse matrix to the matrix representing the metric $h_{ij}=Q^{\ast}{}^A_{i}G^{\rm H}_{AB} Q^{\ast}{}^A_j$     on the base space $\mathcal M$ of the principal fiber bundle $\rm P(\mathcal M,\mathcal G)$.

Next, we transform those terms of the horizontal equation (\ref {itog_hor_1}) that depend on the Christoffel symbols:
\begin{equation}
 N^A_R\,{}^{\rm  H}\tilde {\Gamma}^R_{\tilde B\tilde M}{\omega}^{\tilde B} {\omega}^{\tilde M}\equiv N^A_R\,{}^{\rm  H}\tilde {\Gamma}^R_{ B M}{\omega}^{ B} {\omega}^{ M}+2N^A_R\,{}^{\rm  H}\tilde {\Gamma}^R_{q B}{\omega}^{q} {\omega}^{B}+N^A_R\,{}^{\rm  H}\tilde {\Gamma}^R_{ab}{\omega}^{a} {\omega}^{b}.
\label{hor_christoff_symb}
\end{equation}
(In our notation, the index formed by a capital  letter  with a tilde, for example, $\tilde B$,  means that $\tilde B=(B,b)$, so we have a summation  over two indices.) 

We must express the Christoffel symbols ${}^{\rm  H}\tilde {\Gamma}^R_{\tilde B\tilde M}(Q^{\ast}(x))$ through the Christoffel symbols determined  for the Riemannian metric  on the base space of the principal fiber bundle $\pi': \mathcal P\times  V \to \mathcal P\times_{\mathcal G}  V $. In  local coordinates $(x^i,\tilde f^a),$ this metric is represented by the following matrix:
\begin{equation}
\displaystyle 
\left(
\begin{array}{cc}
\!\!\!\!\!\!\!\!\!\tilde h_{ij} & \tilde G ^{\rm H}_{ia}\,Q^{\ast}{}^A_i\\
\tilde G ^{\rm H}_{Bb}\,Q^{\ast}{}^B_j & \tilde G ^{\rm H}_{ba}
\end{array}
\right),
\label{hor_metric_x}
\end{equation}
where $\tilde h_{ij}=Q^{\ast}{}^A_i \tilde G ^{\rm H}_{AB}    Q^{\ast}{}^B_j$.

We recall that, in our notation, the tilde sign in $\tilde G ^{\rm H}_{AB}, \tilde G ^{\rm H}_{ia}, {}^{\rm  H}\tilde {\Gamma}$  means that the corresponiding quantities are determined using the  metric $d_{\alpha \beta}=\gamma_{\alpha \beta}+\gamma'_{\alpha \beta}$ on the orbits of the bundle $\pi'$.

The Christoffel symbols ${}^{\rm  H}\tilde {\Gamma}^i_{jk}$ for the metric (\ref{hor_metric_x}) are calculated using the following formula:
\begin{equation}
 {}^{\rm  H}\tilde {\Gamma}^i_{jk}=\tilde h^{il}\,{}^{\rm  H}\tilde {\Gamma}_{jkl}+\tilde h^{ia}\,{}^{\rm  H}\tilde {\Gamma}_{jka},
\label{christoff_xij}
\end{equation}
where $\tilde h^{il}$ and $\tilde h^{ia}$ are the components of the matrix which is inverse to the matrix (\ref{hor_metric_x}):
\begin{eqnarray*}
 \displaystyle
\left(
\begin{array}{cc}
 \tilde h^{ji} & \tilde h^{jb}\\
\tilde h^{ci} & \tilde h^{cb}
\end{array}
\right)
\left(
\begin{array}{cc}
 \tilde h_{in} & \tilde h_{ia}\\
\tilde h_{bn} & \tilde h_{ba}
\end{array}
\right)=
\left(
\begin{array}{cc}
 \delta ^j_n & 0\\
0 & \delta ^c_a
\end{array}
\right).
\end{eqnarray*}
The components of the inverse matrix are given by 
\begin{eqnarray*}
 &&\tilde h^{ji}=G^{EF}N^S_EN^D_FT^j_ST^i_D\\
&&\tilde h^{jb}=G^{EF}N^b_FN^P_ET^j_P\\
&&\tilde h^{cb}=G^{cb}+G^{EF}N^c_EN^b_F\,.
\end{eqnarray*}
In these formulae by $T^i_A$ we denote the projection operator defined as
\[
 T^i_A=(P_{\bot})^D_A(Q^{\ast}(x))G^{\rm H}_{DL}(Q^{\ast}(x))Q^{\ast}{}^L_m(x) h^{mi}(x).
\]
It has  the following important properties: $T^i_AQ^{\ast}{}^A_k=\delta^i_k$ and $Q^{\ast}{}^A_iT^i_B=(P_{\bot})^A_B$.
In addition, the operator $T^i_A$ is used to replace the partial derivative with respect to the dependent variable $Q^{\ast}{}^A$: $\frac{\partial}{\partial Q^{\ast}{}^A}=T^i_A\frac{\partial}{\partial x^i}$.

Christoffel symbols ${}^{\rm  H}\tilde {\Gamma}_{ijk}$ and ${}^{\rm  H}\tilde {\Gamma}_{ija}$ with standard definition, such as  ${}^{\rm  H}\tilde {\Gamma}_{ijk}=\frac12(\tilde h_{ik,j}+ \tilde h_{jk,i}-\tilde h_{ij,k})$, lead to the following equalities: 
\begin{eqnarray}
 &&{}^{\rm  H}\tilde {\Gamma}_{ijk}={}^{\rm  H}\tilde {\Gamma}_{ABC}Q^{\ast}{}^A_iQ^{\ast}{}^B_jQ^{\ast}{}^C_k+\tilde G^{\rm H}_{AB}Q^{\ast}{}^A_{ij}Q^{\ast}{}^B_k
\label{gamma_ijk_x}\\
&&{}^{\rm  H}\tilde {\Gamma}_{ija}={}^{\rm  H}\tilde {\Gamma}_{ABa}Q^{\ast}{}^A_iQ^{\ast}{}^B_j+\tilde G^{\rm H}_{Aa}Q^{\ast}{}^A_{ij}.
\label{gamma_ija_x}
\end{eqnarray}

Since ${}^{\rm  H}\tilde {\Gamma}_{ABC}=\tilde G^{\rm H}_{\tilde R C}{}^{\rm  H}\tilde {\Gamma}^{\tilde R}_{AB}$ and 
${}^{\rm  H}\tilde {\Gamma}_{ABa}=\tilde G^{\rm H}_{\tilde R a}{}^{\rm  H}\tilde {\Gamma}^{\tilde R}_{AB}$, the previous equalities can be rewritten as 
\begin{eqnarray*}
 &&{}^{\rm  H}\tilde {\Gamma}_{ijk}=\tilde G^{\rm H}_{\tilde R C}{}^{\rm  H}\tilde {\Gamma}^{\tilde R}_{AB}Q^{\ast}{}^A_iQ^{\ast}{}^B_jQ^{\ast}{}^C_k+\tilde G^{\rm H}_{AB}Q^{\ast}{}^A_{ij}Q^{\ast}{}^B_k\,,\\
&&{}^{\rm  H}\tilde {\Gamma}_{ija}=\tilde G^{\rm H}_{\tilde R a}{}^{\rm  H}\tilde {\Gamma}^{\tilde R}_{AB}Q^{\ast}{}^A_iQ^{\ast}{}^B_j+\tilde G^{\rm H}_{Aa}Q^{\ast}{}^A_{ij}.
\end{eqnarray*}
The obtained expressions for Christoffel symbols are used in the formula (\ref{christoff_xij}).

It follows that the first term on the right-hand side of Eqs.(\ref{christoff_xij}), 
$$G^{EF}N^S_EN^D_FT^i_ST^l_D\Bigl(\tilde G^{\rm H}_{\tilde R C}{}^{\rm  H}\tilde {\Gamma}^{\tilde R}_{AB}Q^{\ast}{}^A_jQ^{\ast}{}^B_kQ^{\ast}{}^C_l+\tilde G^{\rm H}_{AB}Q^{\ast}{}^A_{jk}Q^{\ast}{}^B_l\Bigr),$$
can be rewritten as 
$$G^{EF}N^S_EN^C_FT^i_S\,\tilde G^{\rm H}_{  \tilde R C}{}^{\rm  H}\tilde {\Gamma}^{\tilde R}_{AB}Q^{\ast}{}^A_jQ^{\ast}{}^B_k+G^{EF}N^S_EN^B_FT^i_S\,\tilde G^{\rm H}_{AB}Q^{\ast}{}^A_{jk}.$$
To get this, we used the following properties: $T^l_DQ^{\ast}{}^C_l=(P_{\bot})^C_D$ and $N^D_F(P_{\bot})^C_D=N^C_F$.

The second term on the right-hand side of Eqs.(\ref{christoff_xij}) is equal to
$$G^{EF}N^S_EN^a_FT^i_S\,\tilde G^{\rm H}_{\tilde R a}{}^{\rm  H}\tilde {\Gamma}^{\tilde R}_{AB}Q^{\ast}{}^A_jQ^{\ast}{}^B_k+G^{EF}N^S_EN^a_FT^i_S\,\tilde G^{\rm H}_{Aa}Q^{\ast}{}^A_{jk}.$$

Therefore, the expression on the right side of Eqs.(\ref{christoff_xij}) can be rewritten as
\begin{eqnarray*}
&&G^{EF}N^S_EN^C_FT^i_S\Bigl(N^C_F\tilde G^{\rm H}_{  \tilde R C}
+N^a_F\tilde G^{\rm H}_{  \tilde R a}\Bigr){}^{\rm  H}\tilde {\Gamma}^{\tilde R}_{AB}Q^{\ast}{}^A_jQ^{\ast}{}^B_k\\
&&+G^{EF}N^S_ET^i_S\Bigl(N^B_F \tilde G^{\rm H}_{AB}+N^a_F \tilde G^{\rm H}_{Aa}\Bigr) Q^{\ast}{}^A_{jk}. 
\end{eqnarray*}
The expression in the first bracket is equal to $\tilde G^{\rm H}_{F  \tilde R}$ (with $\tilde R=(R,r)$).  But $G^{EF}\tilde G^{\rm H}_{F   R}=\tilde \Pi^E_R$ and  $ N^S_E \tilde \Pi^E_R=N^S_R$, while  $G^{EF}\tilde G^{\rm H}_{F   r}=\tilde \Pi^E_r$ and  $ N^S_E \tilde \Pi^E_r=0$.

The expression in the second bracket is equal to $\tilde G^{\rm H}_{F  A}$. Now the same arguments lead to the multiplier $N^S_A$, which stands before $Q^{\ast}{}^A_{jk}$.

So, we get 
\[
 {}^{\rm  H}\tilde {\Gamma}^i_{jk}=T^i_SN^S_R\Bigl({}^{\rm  H}\tilde {\Gamma}^{\tilde R}_{AB}Q^{\ast}{}^A_jQ^{\ast}{}^B_k+ Q^{\ast}{}^R_{jk}\Bigr).
\]
Multiplying both part of this equality by $Q^{\ast}{}^D_i$ we come to
\[
 Q^{\ast}{}^D_i\,{}^{\rm  H}\tilde {\Gamma}^i_{jk}=N^D_R\Bigl({}^{\rm  H}\tilde {\Gamma}^{\tilde R}_{AB}Q^{\ast}{}^A_jQ^{\ast}{}^B_k+ Q^{\ast}{}^R_{jk}\Bigr).
\]
It follows that the first term with the Christoffel symbol on the right of Eqs.(\ref{hor_christoff_symb}), together  with the kinetic term, is transformed into
\[
 Q^{\ast}{}^A_i\Bigl(\ddot x^i+{}^{\rm  H}\tilde {\Gamma}^i_{jk}\dot x^j\dot x^k\Bigr).
\]

The second term $N^A_R\,{}^{\rm  H}\tilde {\Gamma}^R_{a B}{\omega}^{a} {\omega}^{B}$    on the right of Eqs.(\ref{hor_christoff_symb}) is related to 
${}^{\rm  H}\tilde {\Gamma}^i_{aj}\dot{\tilde f}^a\dot x^j$, where the Christoffel symbol ${}^{\rm  H}\tilde {\Gamma}^i_{aj}$ for the metric (\ref{hor_metric_x}) is defined  by 
\begin{equation}
 {}^{\rm  H}\tilde {\Gamma}^i_{aj}=\tilde h^{ik}\, {}^{\rm  H}\tilde {\Gamma}_{ajk}+\tilde h^{ib}\,{}^{\rm  H}\tilde {\Gamma}_{ajb}\,.
\label{christoff_xaj}
\end{equation}
 This can be shown as follows.

Calculating ${}^{\rm  H}\tilde {\Gamma}_{ajk}=\frac12(\tilde h_{ak,j}+ \tilde h_{jk,a}-\tilde h_{aj,k})$, we get
\[
 {}^{\rm  H}\tilde {\Gamma}_{ajk}={}^{\rm  H}\tilde {\Gamma}_{aBA}Q^{\ast}{}^B_jQ^{\ast}{}^A_k.
\]
Similarly, for ${}^{\rm  H}\tilde {\Gamma}_{ajb}$ we obtain
\[
 {}^{\rm  H}\tilde {\Gamma}_{ajb}={}^{\rm  H}\tilde {\Gamma}_{aDb}Q^{\ast}{}^D_j.
\]
By definition, 
$${}^{\rm  H}\tilde {\Gamma}_{aBA}=\tilde G^{\rm H}_{\tilde R A}{}^{\rm  H}\tilde {\Gamma}^{\tilde R}_{aB}\;\;\; \rm{and} \;\;\;
{}^{\rm  H}\tilde {\Gamma}_{aDb}=\tilde G^{\rm H}_{\tilde R b}{}^{\rm  H}\tilde {\Gamma}^{\tilde R}_{aD}.$$
Using these expressions in (\ref{christoff_xaj}), together with the explicit representations of $\tilde h^{ik}$ and $\tilde h^{ib}$,
we get
\[
{}^{\rm  H}\tilde {\Gamma}^i_{aj}=G^{EF}N^S_ET^i_S\Bigl(N^A_F\tilde G^{\rm H}_{  \tilde R A}
+N^b_F\tilde G^{\rm H}_{  \tilde R b}\Bigr){}^{\rm  H}\tilde {\Gamma}^{\tilde R}_{aB}Q^{\ast}{}^B_j.
\]
It is not difficult to see that $N^A_F\tilde G^{\rm H}_{  \tilde R A}+N^b_F\tilde G^{\rm H}_{  \tilde R b}=\tilde G^{\rm H}_{  \tilde R F}$, so we can  do  the same transformations as in our previous calculation. As a result, we obtain:
\[
 {}^{\rm  H}\tilde {\Gamma}^i_{aj}=T^i_SN^S_R\,{}^{\rm  H}\tilde {\Gamma}^{R}_{aB}Q^{\ast}{}^B_j.
\]
Multiplying the obtained equality by $Q^{\ast}{}^A_i$, we get
\[
 Q^{\ast}{}^A_i\,{}^{\rm  H}\tilde {\Gamma}^i_{aj}=N^A_R\,{}^{\rm  H}\tilde {\Gamma}^{R}_{aB}Q^{\ast}{}^B_j.
\]
Thus, the second term on the right of Eqs.(\ref{hor_christoff_symb}) is  transformed as follows:
$$N^A_R\,{}^{\rm  H}\tilde {\Gamma}^R_{a B}{\omega}^{a} {\omega}^{B}=Q^{\ast}{}^A_i\,{}^{\rm  H}\tilde {\Gamma}^i_{aj}\dot{\tilde f}^a\dot x^j.$$ 

The third term $N^A_R\,{}^{\rm  H}\tilde {\Gamma}^R_{ab}{\omega}^{a} {\omega}^{b}$ on the right of Eqs.(\ref{hor_christoff_symb}) can be transformed in a similar way. Here we have
\[
 {}^{\rm  H}\tilde {\Gamma}^i_{ab}=\tilde h^{ik}\, {}^{\rm  H}\tilde {\Gamma}_{abk}+\tilde h^{ic}\,{}^{\rm  H}\tilde {\Gamma}_{abc},
\]
${}^{\rm  H}\tilde {\Gamma}_{ab\tilde T}=\tilde G^{\rm H}_{\tilde R \tilde T}{}^{\rm  H}\tilde {\Gamma}^{\tilde R}_{ab}$. And we get
$$N^A_R\,{}^{\rm  H}\tilde {\Gamma}^R_{a b}{\omega}^{a} {\omega}^{b}=Q^{\ast}{}^A_i\,{}^{\rm  H}\tilde {\Gamma}^i_{ab}\dot{\tilde f}^a\dot{\tilde f}^b.$$


Next we consider the transformation of $\mathscr F$-terms of the horizontal equation (\ref{itog_hor_1}).
These terms  are given by
\begin{eqnarray*}
&&G^{EF}N^A_E\bigl(N^R_F{\mathscr F}^{\alpha}_{\tilde B R}p_{\alpha}+N^r_F{\mathscr F}^{\alpha}_{\tilde B r}\bigr)\omega^{\tilde B}p_{\alpha}=\nonumber\\
&&G^{EF}N^A_E(N^R_F{\mathscr F}^{\alpha}_{B R}+N^r_F{\mathscr F}^{\alpha}_{ B r})\omega^{ B}p_{\alpha}+
G^{EF}N^A_E(N^R_F{\mathscr F}^{\alpha}_{b R}+N^r_F{\mathscr F}^{\alpha}_{b r})\omega^{b}p_{\alpha},
\end{eqnarray*}
where $ {\mathscr F}^{\alpha}_{BR} $ is defined as
\[
{\mathscr F}^{\alpha}_{BR}=\displaystyle\frac{\partial}{\partial Q^{\ast}{}^B}\,{\mathscr A}^{\alpha}_R- 
\frac{\partial}{\partial {Q^{\ast}}^R}\,{\mathscr A}^{\alpha}_B
+c^{\alpha}_{\nu\sigma}\, {\mathscr A}^{\nu}_B\,
{\mathscr A}^{\sigma}_R.
\]

First we will study the transformation of the following expression ${\mathscr F}^{\alpha}_{BR}Q^{\ast B}_kQ^{\ast R}_l$, in which one of the multiplier, $Q^{\ast B}_k$, occurs due to  $\omega ^B=Q^{\ast B}_k\dot x^k$.

Since   $\frac{\partial}{\partial Q^{\ast}{}^B}=T^j_B\frac{\partial}{\partial x^j}$, the first term in ${\mathscr F}^{\alpha}_{BR}Q^{\ast B}_kQ^{\ast R}_l$ can be rewritten as follows:
\begin{eqnarray*}
&&\biggl(\frac{\partial}{\partial Q^{\ast}{}^B}\,{\mathscr A}^{\alpha}_R\biggr)Q^{\ast B}_kQ^{\ast R}_l=T^j_B\biggl(\frac{\partial}{\partial x^j}\,{\mathscr A}^{\alpha}_R\biggr)Q^{\ast B}_kQ^{\ast R}_l=
\nonumber\\
&&\frac{\partial}{\partial x^k}\biggl(\,{\mathscr A}^{\alpha}_RQ^{\ast R}_l\biggr)-{\mathscr A}^{\alpha}_RQ^{\ast R}_{lk}\equiv
\frac{\partial}{\partial x^k}{\mathscr A}^{\alpha}_l-{\mathscr A}^{\alpha}_RQ^{\ast R}_{lk}.
\end{eqnarray*}
A similar representation can also be obtained for the second term in ${\mathscr F}^{\alpha}_{BR}Q^{\ast B}_kQ^{\ast R}_l$.
The sum of the first and second  terms together with the third term leads to the following equality: 
\[
 {\mathscr F}^{\alpha}_{BR}Q^{\ast B}_kQ^{\ast R}_l={\mathscr F}^{\alpha}_{kl}.
\]
Using this equality, it can be shown that 
\[
 Q^{\ast L}_m \tilde h^{ml}{\mathscr F}^{\alpha}_{kl}=G^{EF}N^L_EN^R_F{\mathscr F}^{\alpha}_{B R}Q^{\ast B}_k.
\]

In the second $\mathscr F$-term of the horizontal equation (\ref{itog_hor_1}), we first need to transform the expression 
${\mathscr F}^{\alpha}_{Br}Q^{\ast B}_k$, in which
\[
{\mathscr F}^{\alpha}_{Br}=\displaystyle\frac{\partial}{\partial Q^{\ast}{}^B}\,{\mathscr A}^{\alpha}_r- 
\frac{\partial}{\partial \tilde f^r}\,{\mathscr A}^{\alpha}_B
+c^{\alpha}_{\nu\sigma}\, {\mathscr A}^{\nu}_B\,
{\mathscr A}^{\sigma}_r,
\]
and ${\mathscr A}^{\alpha}_r=d^{\alpha\mu}K^p_{\mu}G_{pr}$ with $d^{\alpha\mu}=d^{\alpha\mu}(Q^{\ast},\tilde f)$.
Then, acting in the same way as above, we get that
\[
 {\mathscr F}^{\alpha}_{Br}Q^{\ast B}_k={\mathscr F}^{\alpha}_{kr}.
\]
It can be verified that in this case the following relationship is obtained:
\[
 G^{EF}N^A_EN^a_F{\mathscr F}^{\alpha}_{B a}Q^{\ast B}_k=Q^{\ast A}_m \tilde h^{ma}{\mathscr F}^{\alpha}_{ka}.
\]

For the third $\mathscr F$-term,  which is given by $G^{EF}N^A_EN^R_F{\mathscr F}^{\alpha}_{b R}\omega^{b}$, 
\[
{\mathscr F}^{\alpha}_{bR}=\displaystyle\frac{\partial}{\partial \tilde f^b}\,{\mathscr A}^{\alpha}_R- 
\frac{\partial}{\partial Q^{\ast}{}^R}\,{\mathscr A}^{\alpha}_b
+c^{\alpha}_{\nu\sigma}\, {\mathscr A}^{\nu}_b\,
{\mathscr A}^{\sigma}_R\equiv T^i_R{\mathscr F}^{\alpha}_{bi},
\]
one can obtain the following relation:
\[
 G^{EF}N^A_EN^R_F{\mathscr F}^{\alpha}_{b R}\omega^{b}=Q^{\ast A}_m \tilde h^{mi}{\mathscr F}^{\alpha}_{bi}\dot{\tilde f ^b}.
\]

And for the last $\mathscr F$-term in (\ref{itog_hor_1}), we get the following:
\[
 G^{EF}N^A_EN^a_F{\mathscr F}^{\alpha}_{b a}\omega^{b}=Q^{\ast A}_m \tilde h^{ma}{\mathscr F}^{\alpha}_{ba}\dot{\tilde f ^b}.
\]

 In the first horizontal equation, the terms with covariant derivatives are given by two expressions:
\[
G^{EF}N^A_EN^{\tilde R}_F(\mathscr D_{\tilde R} d^{\kappa \sigma})p_{\kappa}p_{\sigma}= \Bigl(G^{EF}N^A_EN^R_F\mathscr D_R d^{\kappa \sigma}+G^{EF}N^A_EN^a_F\mathscr D_a d^{\kappa \sigma}\Bigr)p_{\kappa}p_{\sigma},
\]
where
\[
 \mathscr D_R d^{\kappa \sigma}=\displaystyle\frac{\partial}{\partial Q^{\ast}{}^R}d^{\kappa \sigma}+c^{\kappa}_{\mu\nu}{\mathscr A}^{\mu}_Rd^{\nu \sigma}+c^{\sigma}_{\mu\nu}{\mathscr A}^{\mu}_Rd^{\nu \kappa}
\]
and
\[
 \mathscr D_a d^{\kappa \sigma}=\displaystyle\frac{\partial}{\partial \tilde f^a}d^{\kappa \sigma}+c^{\kappa}_{\mu\nu}{\mathscr A}^{\mu}_ad^{\nu \sigma}+c^{\sigma}_{\mu\nu}{\mathscr A}^{\mu}_ad^{\nu \kappa}.
\]
Replacing the dependent  coordinates $Q^{\ast A}$ with the functions $Q^{\ast A}(x)$  and using the properties of the projectors $N^A_B$, $(P_{\bot})^A_B$ and $T^i_D$ when transforming $Q^{\ast A}$ to $x^i$, allows us to represent terms with covariant derivatives as follows:
\[
 G^{EF}N^A_EN^{\tilde R}_F(\mathscr D_{\tilde R} d^{\kappa \sigma})p_{\kappa}p_{\sigma}=Q^{\ast A}_m(\tilde h^{mi}\mathscr D_i d^{\kappa \sigma}+\tilde h^{ma}\mathscr D_a d^{\kappa \sigma})p_{\kappa}p_{\sigma}.
\]

A similar approach also leads us to the representation of a potential term in the first horizontal equation:
\[
G^{EF}N^A_EN^{\tilde R}_F V_{,\tilde R}=Q^{\ast A}_m \displaystyle \biggl(\tilde h^{mi}\frac{\partial}{\partial x^i}V+\tilde h^{ma}\frac{\partial}{\partial \tilde f^a}V\biggr).
\]

Thus, as a result of all these transformations performed in equation (\ref{itog_hor_1}), we get the following equation:
\begin{eqnarray}
&&Q^{\ast}{}^A_i\Bigl(\ddot x^i+{}^{\rm  H}\tilde {\Gamma}^i_{jk}\dot x^j\dot x^k+2\,{}^{\rm  H}\tilde {\Gamma}^i_{aj}\dot{\tilde f}^a\dot x^j+{}^{\rm  H}\tilde {\Gamma}^i_{ab}\dot{\tilde f}^a\dot{\tilde f}^b \Bigr.
\nonumber\\ 
&&+(\tilde h^{il}{\mathscr F}^{\alpha}_{kl}+\tilde h^{ia}{\mathscr F}^{\alpha}_{ka})\dot x^kp_{\alpha}
+(\tilde h^{in}{\mathscr F}^{\alpha}_{bn}+\tilde h^{ia}{\mathscr F}^{\alpha}_{ba})\dot{\tilde f ^b}p_{\alpha}
\nonumber\\
&&+\Bigl.\frac12(\tilde h^{in}\mathscr D_n d^{\kappa \sigma}+\tilde h^{ia}\mathscr D_a d^{\kappa \sigma})p_{\kappa}p_{\sigma}
+\displaystyle \biggl(\tilde h^{in}\frac{\partial}{\partial x^n}V+\tilde h^{ia}\frac{\partial}{\partial \tilde f^a}V\biggr)\Bigr)=0.
\label{itog_hor_1_x}
\end{eqnarray}

Now consider the transformation of the second horizontal equation (\ref{itog_hor_2}).
The kinetic term of this equation is $ N^b_B\frac{d \omega^B}{dt}+\frac{d\omega ^b}{dr}$, where
$\omega^B\equiv \frac{d Q^{\ast B}(x(t))}{dt}\equiv Q^{\ast B}_i \dot x^i$ and 
$\omega ^b=\dot{\tilde f^b}$. Since
\[
 \frac{d}{dt}\Bigl( Q^{\ast B}_i\dot x^i\Bigr)= Q^{\ast B}_i\ddot x^i+ Q^{\ast B}_{ij}\dot x^i\dot x^j
\]
and $N^b_B Q^{\ast B}_i=0$, we get $N^b_B\, Q^{\ast B}_{ij}\dot x^i\dot x^j+
\ddot{\tilde f^b}$.

The terms of  (\ref{itog_hor_2}) with the Christoffel symbols are given by
\begin{equation}
 N^b_{\tilde R}\Bigl({}^{\rm  H}\tilde {\Gamma}^{\tilde R}_{AB}Q^{\ast A}_iQ^{\ast B}_j\dot x^i\dot x^j+2\,{}^{\rm  H}\tilde {\Gamma}^{\tilde R}_{Aa}Q^{\ast A}_i\dot x^i\dot{\tilde f^a}+{}^{\rm  H}\tilde {\Gamma}^{\tilde R}_{ac}\dot {\tilde f^a}\dot {\tilde f^c}\Bigr).
\label{christ_symbl_2_hor_eq}
\end{equation}
First we will deal with the transformation of the $(\dot x^i\dot x^j)$-term. It is natural to assume that
$$N^b_{\tilde R}{}^{\rm  H}\tilde {\Gamma}^{\tilde R}_{AB}Q^{\ast A}_iQ^{\ast B}_j\equiv 
N^b_{R}{}^{\rm  H}\tilde {\Gamma}^{R}_{AB}Q^{\ast A}_iQ^{\ast B}_j+N^b_{c}\,{}^{\rm  H}\tilde {\Gamma}^{c}_{AB}Q^{\ast A}_iQ^{\ast B}_j$$ can be expressed through 
${}{}^{\rm  H}\tilde {\Gamma}^{b}_{ij}$. This can be verified as follows.

By definition,
\[
 {}{}^{\rm  H}\tilde {\Gamma}^{b}_{ij}=\tilde h^{bk}\,{}^{\rm  H}\tilde {\Gamma}_{ijk}+\tilde h^{ba}\,{}^{\rm  H}\tilde {\Gamma}_{ija},
\]
with $\tilde h^{bk}=G^{EF}N^b_FN^P_ET^k_P$ and $\tilde h^{ba}=G^{ba}+G^{EF}N^b_FN^a_E$.
Using (\ref{gamma_ijk_x}) and (\ref{gamma_ija_x}) for ${}^{\rm  H}\tilde {\Gamma}_{ijk}$ and ${}^{\rm  H}\tilde {\Gamma}_{ija}$, we get
\begin{eqnarray*}
 &&{}{}^{\rm  H}\tilde {\Gamma}^{b}_{ij}=G^{EF}N^b_FN^P_ET^k_P\Bigl(\tilde G^{\rm H}_{\tilde R C}{}^{\rm  H}\tilde {\Gamma}^{\tilde R}_{AB}Q^{\ast A}_iQ^{\ast B}_jQ^{\ast C}_k
+\tilde G^{\rm H}_{AB}Q^{\ast A}_{ij}Q^{\ast B}_k\Bigr)\nonumber\\
&&+\Bigl(G^{ba}+G^{EF}N^b_FN^a_E\Bigr)\Bigl(\tilde G^{\rm H}_{\tilde R a}{}^{\rm  H}\tilde {\Gamma}^{\tilde R}_{AB}Q^{\ast A}_iQ^{\ast B}_jQ^{\ast C}_k
+\tilde G^{\rm H}_{Aa}Q^{\ast A}_{ij}\Bigr).
\nonumber\\
\end{eqnarray*}
Because of $T^k_PQ^{\ast C}_k=(P_{\bot})^C_P$ and $N^P_E(P_{\bot})^C_P=N^C_E$, the right-hand side of the above equation is rewritten as 
\begin{eqnarray*}
 &&G^{EF}N^b_FN^C_E\tilde G^{\rm H}_{\tilde R C}{}^{\rm  H}\tilde {\Gamma}^{\tilde R}_{AB}Q^{\ast A}_iQ^{\ast B}_j+G^{EF}N^b_FN^B_E\tilde G^{\rm H}_{AB}Q^{\ast A}_{ij}
\nonumber\\
&&+\Bigl(G^{ba}+G^{EF}N^b_FN^a_E\Bigr)\Bigl(\tilde G^{\rm H}_{\tilde R a}{}^{\rm  H}\tilde {\Gamma}^{\tilde R}_{AB}Q^{\ast A}_iQ^{\ast B}_jQ^{\ast C}_k
+\tilde G^{\rm H}_{Aa}Q^{\ast A}_{ij}\Bigr).
\nonumber\\
\end{eqnarray*}
 Consider the terms of the previous expression that depend on $ \tilde {\Gamma}^{\tilde R}_{AB}$. That is, the following terms:
\[
 G^{EF}N^b_F\Bigl(N^C_E\tilde G^{\rm H}_{\tilde R C}+N^a_E\tilde G^{\rm H}_{\tilde R a}\Bigr)\tilde {\Gamma}^{\tilde R}_{AB}Q^{\ast A}_iQ^{\ast B}_j+G^{ba}\tilde G^{\rm H}_{\tilde Ra}\tilde {\Gamma}^{\tilde R}_{AB}Q^{\ast A}_iQ^{\ast B}_j.
\]
It can be shown that the expression  in brackets is equal to ${\tilde G}^{\rm H}_{E\tilde R}$. In addition, we have two identities by which $G^{EF}\tilde G^{\rm H}_{E\tilde R}=\tilde \Pi^F_{\tilde R}$ and $G^{ba}\tilde G^{\rm H}_{\tilde Ra}=\tilde {\Pi}^b_{\tilde R}$. Furthermore, our projectors satisfy the property $N^b_F\tilde {\Pi}^F_{\tilde R}+\tilde{\Pi}^b_{\tilde R}=N^b_{\tilde R}$. Taking into account all these facts,  we  obtain  that the terms with $\tilde {\Gamma}^{\tilde R}_{AB}$ are rewritten as $N^b_{\tilde R}\tilde {\Gamma}^{\tilde R}_{AB} Q^{\ast A}_iQ^{\ast B}_j$.

The terms with $Q^{\ast A}_{ij}$ in the above  representation for ${}^{\rm  H}\tilde {\Gamma}^{b}_{ij}$ can be treated in a similar way. As a result, we come to
\[
 {}^{\rm  H}\tilde {\Gamma}^{b}_{ij}= N^b_{\tilde R}\tilde {\Gamma}^{\tilde R}_{AB}Q^{\ast A}_iQ^{\ast B}_j+N^b_AQ^{\ast A}_{ij}.
\]

Notice that after substituting   this expression into the second horizontal equation (\ref{itog_hor_2}), the ``$Q^{\ast A}_{ij}$-term''   will be mutually canceled by a similar term resulting from the transformation of  the kinetic term of the equation.

The second term in (\ref{christ_symbl_2_hor_eq}), $N^b_{\tilde R}{}^{\rm  H}\tilde {\Gamma}^{\tilde R}_{Aa}Q^{\ast A}_i\dot x^i\dot{\tilde f^a}$, is related to ${}^{\rm  H}\tilde {\Gamma}^{b}_{ia}$. This can be shown by using the following representation:
\[
 {}^{\rm  H}\tilde {\Gamma}^{b}_{ia}=\tilde h^{bj}{}^{\rm  H}\tilde {\Gamma}_{iaj}+\tilde h^{bc}{}^{\rm  H}\tilde {\Gamma}_{iac},
\]
 where ${}^{\rm  H}\tilde {\Gamma}_{iaj}={}^{\rm  H}\tilde {\Gamma}_{AaB}Q^{\ast A}_iQ^{\ast B}_j$ and ${}^{\rm  H}\tilde {\Gamma}_{iac}={}^{\rm  H}\tilde {\Gamma}_{Aac}Q^{\ast A}_i$.
Note that now, by definition, we have
${}^{\rm  H}\tilde {\Gamma}_{AaB}=\tilde G^{\rm H}_{\tilde R B}{}^{\rm  H}\tilde {\Gamma}_{Aa}^{\tilde R}$ and  ${}^{\rm  H}\tilde {\Gamma}_{Aac}=\tilde G^{\rm H}_{\tilde R c}{}^{\rm  H}\tilde {\Gamma}_{Aa}^{\tilde R}$.

Performing the same transformation as in the previous case, we get
\[
 {}^{\rm  H}\tilde {\Gamma}^{b}_{ia}=\Bigl(N^b_R{}^{\rm  H}\tilde {\Gamma}_{Aa}^{ R}+{}^{\rm  H}\tilde {\Gamma}_{Aa}^{ b}\Bigr)Q^{\ast A}_i\equiv N^b_{\tilde R}{}^{\rm  H}\tilde {\Gamma}_{Aa}^{ \tilde R}Q^{\ast A}_i.
\]
So, the second term in  (\ref{christ_symbl_2_hor_eq}) is equal to $2\,{}^{\rm  H}\tilde {\Gamma}^{b}_{ia}\dot x^i\dot{\tilde f^a}$.

The last term in (\ref{christ_symbl_2_hor_eq}), $N^b_{\tilde R}{}^{\rm  H}\tilde {\Gamma}^{\tilde R}_{ac}\dot{\tilde f^a}\dot{\tilde f^c}$, is transformed into ${}^{\rm  H}\tilde {\Gamma}^{b}_{ac}\dot{\tilde f^a}\dot{\tilde f^c}$. To show this one must use  the representation
\[
 {}^{\rm  H}\tilde {\Gamma}^{b}_{ac}=\tilde h^{bi}\,{}^{\rm  H}\tilde {\Gamma}_{aci}+\tilde h^{bd}\,{}^{\rm  H}\tilde {\Gamma}_{acd},
\]
where ${}^{\rm  H}\tilde {\Gamma}_{aci}={}^{\rm  H}\tilde {\Gamma}_{acA}\,Q^{\ast A}_i$, ${}^{\rm  H}\tilde {\Gamma}_{acA}=     \tilde G^{\rm H}_{\tilde R A}{}^{\rm  H}\tilde {\Gamma}_{ac}^{\tilde R}$ and ${}^{\rm  H}\tilde {\Gamma}_{acd}=G^{\rm H}_{\tilde R d}{}^{\rm  H}\tilde {\Gamma}_{ac}^{\tilde R}$. 

Note also that the expressions arising in the process of transformation  have been simplified using the identity
 $N^a_F\tilde{\Pi}^F_{\tilde R}+ \tilde{\Pi}^a_{\tilde R}=N^a_{\tilde R}$.

 $\mathscr F$-terms of the second horizontal equation
 (\ref{itog_hor_2}) are given by
\[
\bigl(G^{EF}N^b_FN^{\tilde R}_E{\mathscr F}^{\alpha}_{\tilde B \tilde R}+G^{ba}{\mathscr F}^{\alpha}_{\tilde B a}\bigr)\omega^{\tilde B}p_{\alpha}.
\]
They can be rewritten as follows:
\begin{eqnarray*}
&& \Bigl[G^{EF}N^b_FN^{ R}_E{\mathscr F}^{\alpha}_{B R}Q^{\ast B}_i \dot x^i+G^{EF}N^b_FN^{ R}_E{\mathscr F}^{\alpha}_{d R}\dot{\tilde f^d}\Bigr.\nonumber\\
&&\Bigl.+(G^{ba}+G^{EF}N^b_fN^a_E){\mathscr F}^{\alpha}_{B a}Q^{\ast B}_i \dot x^i+(G^{ba}+G^{EF}N^b_fN^a_E){\mathscr F}^{\alpha}_{d a}\dot{\tilde f^d}\Bigr]p_{\alpha}.
\end{eqnarray*}
Since $G^{EF}N^b_FN^{ R}_E=\tilde h^{jb}Q^{\ast R}_j$ and $Q^{\ast R}_j{\mathscr F}^{\alpha}_{B R}Q^{\ast B}_i={\mathscr F}^{\alpha}_{ij}$, the first term in the above expression is equal to $\tilde h^{jb}{\mathscr F}^{\alpha}_{ij}\dot x^ip_{\alpha}$.
The second term will be equal to $\tilde h^{jb}{\mathscr F}^{\alpha}_{dj}\dot{\tilde f^d}p_{\alpha}$ due to ${\mathscr F}^{\alpha}_{d R}=T^k_R{\mathscr F}^{\alpha}_{d k}$ and $Q^{\ast R}_jT^k_R=\delta ^k_j$. The third term of the considered expression is rewritten as $\tilde h^{ba}{\mathscr F}^{\alpha}_{i a}\dot x^ip_{\alpha}$. And the last term is equal to $\tilde h^{ba}{\mathscr F}^{\alpha}_{d a}\dot{\tilde f^d}p_{\alpha}$. 

The terms with the covariant derivatives and the potential terms of the second horizontal equation (\ref{itog_hor_2}) are transformed in the same way as the corresponding terms of the first horizontal equation. 

So, we get the following representation of the equation (\ref{itog_hor_2}) in  variables ($x^i,{\tilde f^d},p_{\alpha} $):
\begin{eqnarray}
 &&\ddot{\tilde f^b}+{}{}^{\rm  H}\tilde {\Gamma}^{b}_{ij}\dot x^i\dot x^j+2{}^{\rm  H}\tilde {\Gamma}^{b}_{ia}\dot x^i\dot{\tilde f^a}+{}^{\rm  H}\tilde {\Gamma}^{b}_{ac}\dot{\tilde f^a}\dot{\tilde f^c}\nonumber\\
&&+ (\tilde h^{jb}{\mathscr F}^{\alpha}_{ij}+\tilde h^{ba}{\mathscr F}^{\alpha}_{i a})\dot x^ip_{\alpha}
+(\tilde h^{jb}{\mathscr F}^{\alpha}_{dj}+\tilde h^{ba}{\mathscr F}^{\alpha}_{d a})\dot{\tilde f^d}p_{\alpha}\nonumber\\
&&+\frac12\bigl(\tilde h^{ib}\mathscr D_id^{\kappa \sigma}+\tilde h^{ba}\mathscr D_ad^{\kappa \sigma}\bigr)p_{\kappa}p_{\sigma}+
\displaystyle \Bigl(\tilde h^{ib}\frac{\partial}{\partial x^i}V+\tilde h^{ba}\frac{\partial}{\partial \tilde f^a}V\Bigr)=0.
\label{itog_hor_2_x}
\end{eqnarray}
The vertical Lagrange-Poincar\'{e} equation is 
\begin{equation}
 \frac{d p_{\beta}}{dt}+ c^{\nu}_{\mu \beta}d^{\mu \sigma}p_{\sigma}p_{\nu}-c^{\nu}_{\sigma \beta}(\mathscr A^{\sigma}_i\dot x^i+\mathscr A^{\sigma}_a\dot{\tilde f^a})p_{\nu}=0.  
\label{itog_vert_x}
\end{equation}
These equations, together with the horizontal equation (\ref{itog_hor_1_x}), are the transformed Lagrange – Poincaré equations.


\begin{thebibliography}{**}

\bibitem{Stor_1} 
S. N. Storchak. The Lagrange-Poincaré equations for a mechanical system with symmetry on the principal fiber bundle over the base represented by the bundle space of the associated bundle. arXiv:1612.08897 [math-ph]

\bibitem{Stor_2}
S. N. Storchak. Coordinate representation of the Lagrange-Poincaré equations for a mechanical system with symmetry on the total space of a principal fiber bundle whose base is the bundle space of the associated bundle. arXiv:1709.09030  [math-ph]  

\end{thebibliography}
\end{document}